\documentclass[aps,prb,twocolumn,showpacs,superscriptaddress,floatfix,nofootinbib]{revtex4-1}

\usepackage{xcolor} \usepackage[pdftex]{graphicx} \usepackage{textcomp}
\usepackage{amsmath} \usepackage{epsfig} \usepackage{helvet}
\usepackage{amssymb}

 \newcommand{\be}{\begin{equation}}
  \newcommand{\ee}{\end{equation}} \newcommand{\bea}{\begin{eqnarray}}
  \newcommand{\eea}{\end{eqnarray}} \newcommand{\al}{\alpha}
\newcommand{\bet}{\beta}

\newcommand{\tr}{\mbox{tr}} \newcommand{\nn}{\nonumber}
\newcommand{\bra}[1]{\mbox{$\langle #1 |$}}
\newcommand{\ket}[1]{\mbox{$| #1 \rangle$}}
\newcommand{\braket}[2]{\mbox{$\langle #1 | #2 \rangle$}}

\DeclareGraphicsExtensions{.pdf,.png,.jpg}

     \def\tr{ \mbox{tr}} 

\begin{document}
 
\title{Infinite boundary conditions for matrix product state
  calculations}

\author{Ho N. Phien} \affiliation{Centre for Engineered Quantum
  Systems, School of Mathematics and Physics, University of
  Queensland, Brisbane 4072, Australia}

\author{Guifr\'e Vidal} \affiliation{Perimeter Institute for
  Theoretical Physics, Waterloo, Ontario, N2L 2Y5, Canada}

\author{Ian P. McCulloch} \affiliation{Centre for Engineered Quantum
  Systems, School of Mathematics and Physics, University of
  Queensland, Brisbane 4072, Australia}

\date{3/7/2012}

\begin{abstract}
  We propose a formalism to study dynamical properties of a quantum
  many-body system in the thermodynamic limit by studying a finite
  system with \textquotedblleft infinite boundary
  conditions\textquotedblright where both finite-size effects and
  boundary effects have been eliminated. For one-dimensional systems,
  infinite boundary conditions are obtained by attaching two boundary
  sites to a finite system, where each of these two sites effectively
  represents a semi-infinite extension of the system. One can then use
  standard finite-size matrix product state techniques to study a
  region of the system while avoiding many of the complications
  normally associated with finite-size calculations such as boundary
  Friedel oscillations. We illustrate the technique with an example of
  time evolution of a local perturbation applied to an infinite
  (translationally invariant) ground state, and use this to calculate
  the spectral function of the $S=1$ Heisenberg spin chain. This
  approach is more efficient and more accurate than conventional
  simulations based on finite-size matrix product state and
  density-matrix renormalization-group approaches.
\end{abstract}
\pacs{03.67.-a, 03.65.Ud, 02.70.-c, 05.30.Fk}

\maketitle

\tableofcontents
\section{Introduction}
In recent decades, the tensor network formalism has emerged as a set
of powerful numerical techniques to investigate physical properties of
strongly correlated quantum many-body systems. For instance, in 1D
systems, the density-matrix renormalization group (DMRG)\cite{White1,
  White2} is probably the single most powerful method to compute
numerically exact ground states. Furthermore, the development of the
time-evolving block decimation (TEBD) algorithm\cite{Vidal1,Vidal2}
highlighted the great advantages of the matrix product state
(MPS)\cite{Ostlund1, Fannes1, Perez1} representation, which
incorporates DMRG and TEBD into the same framework.\cite{White3,
  Daley1} Meanwhile, tensor product state
(TPS)\cite{Nishino1,Nishino2,Gendiar1,Maeshima1,Nishio1,Gendiar2} and
projected entangled-pair state (PEPS)\cite{Verstraete1,Murg1,Jordan1}
methods are developing into important tools for the study of 2D
systems.

For calculating bulk properties of matter, it is desirable to take the
thermodynamic limit and avoid the influence of boundary conditions. In
many methods, the thermodynamic limit is not possible to study
directly, but instead requires the extrapolation of results for
increasingly larger system sizes. This is because for most algorithms
the computational cost increases with the system size; however approaching
the thermodynamic limit in this way is computationally expensive. 
In 1D, there exist algorithms that overcome this limit by taking advantage of the invariance under
translation in space. One of these is the infinite time-evolving block
decimation (iTEBD),\cite{Vidal3,Roman1} originally introduced to investigate
the time evolution problem for infinite-size 1D spin chains. In this
algorithm, the infinite MPS (iMPS) is represented by a small set of
tensors which are invariant under translation of one unit cell (equal
to two sites for the usual TEBD scheme). This algorithm can be used to
obtain a translationally invariant ground state by evolving the
tensors in imaginary time until the fixed point is reached. The
resulting iMPS is not only a good representation of ground state, but
compared with finite MPS the number of wave function parameters is
reduced and the iMPS form is very convenient for calculating
observables of the system in the thermodynamic limit. The iTEBD
algorithm is very easy to implement; however there are many ways to
optimize an iMPS to achieve the same fixed point. A faster converging algorithm
which also allows more flexibility in the size of the unit cell
is the iDMRG\cite{Ian, Greg} algorithm, but other algorithms exist
with some advantages for some situations.\cite{FrankEvo,SandvikMPS}

Although the iMPS representation of a wave function is very useful for
studying physical systems in the thermodynamic limit, there are some
applications for which breaking of translational invariance is
essential, such as the response to a local perturbation. The time
evolution of a local perturbation is a common technique used in MPS
calculations to obtain the spectral function\cite{White3, White4}
which to date has required using a finite MPS representation.
However, the use of a finite MPS has several disadvantages.  In
particular, the system size needs to be large enough that the
excitation is not influenced by the boundary of the system. This
clearly requires that the propagating excitation will not hit the
boundary, but even this is not enough since the boundary will induce
inhomogeneities such as Friedel oscillations, which means that the
system size must be quite large even to obtain an approximately
homogeneous ground state in the central region of the lattice.

The notion of translational invariance of an iMPS can be generalized
to states with finite momentum, whereby instead of requiring
invariance under some number of lattice shifts, we instead require
only that the iMPS is an eigenstate of translations with some complex
eigenvalue $e^{ik}$ representing the momentum. The resulting iMPS
remains \textit{position independent} but is constructed in such a way
that the transfer operator has non-trivial phase factors. Algorithms
have been proposed for expectation values\cite{Ostlund1,Ian1} and
quasi-particle excitations\cite{Haegeman} using this scheme.

For infinite-size systems an equivalent problem was also investigated
in Ref. \onlinecite{Banuls1}. In that work the authors proposed an
efficient method to simulate both imaginary- and real-time evolution of
the infinite-size system with impurities by transversely contracting
the tensor networks along the space direction rather than along the
time direction as in standard iTEBD. By using a folding technique to reduce
the entanglement of the MPS representation the transverse contraction approach
can achieve longer times than other techniques, nevertheless it
cannot avoid some drawbacks. For instance, by
employing the Suzuki-Trotter decomposition\cite{Suzu1} in the
evolution operator with small time step, the finite number of
rows along the time axis may be very large. This may cause difficulty in
finding the left and right dominant eigenvectors of the transfer
matrix.

We will investigate the above problem in a different way by
introducing what we call {\it infinite boundary conditions} for a
finite MPS. We begin with the ground state of a many-body 1D system
described by an iMPS. A finite region of the
infinite system can be perturbed while still utilizing the iMPS
structure for the tensors not directly affected by the perturbation.
The resulting structure is equivalent to a finite MPS with a specially
constructed `pseudosite' at each end which effectively represents an
infinite extension of the system. A key point of this construction is
that the Hilbert space for the infinite extension is fixed but the
wave function is not, hence it can freely explore all of the available
states in the effective Hilbert space of the infinite extension. The
result is that, in contrast to conventional finite-size MPS
calculations where a propagating excitation reaching the boundary of
the system will reflect back, in our scheme an excitation can
propagate off the end of the finite MPS. As long as the
perturbation outside the finite boundary
is not too big there is little loss in fidelity from
allowing it to do so. 

The evolution of a finite section of an iMPS was considered by Kj\"all \textit{et al.}, \cite{KjallMPS} who used this
notion in obtaining
the time evolution of a translationally invariant state that had been
perturbed by a local particle excitation. However their scheme
was rather specific to the particular setting, of Suzuki-Trotter-based
real-time evolution. In this paper we show that
this idea can be taken much farther, and by mapping the problem onto
a finite MPS then \emph{any} algorithm for finite MPS
calculations can be applied to an infinite system.

The paper is organized as follows: In Sec. II we will introduce the
infinite boundary condition definition and effective Hamiltonian
calculation. In Sec. III we review the problem of a local perturbation
in the infinite spin chain and real-time evolution algorithm. 
In Sec. IV we then apply the idea of infinite boundary conditions to simulate the time
evolution of the spin-1 isotropic antiferromagnetic Heisenberg model. 
The results are presented by calculating time-dependent
observables such as local magnetization $\langle S_{z}(x,t)\rangle$ to see how
a wave front propagates in time and unequal-time two-point correlator
$A(x,t)$, from which we can extract the spectral function and
dispersion relation of the system. Finally, Sec. V contains our conclusions.

\section{Infinite boundary conditions}
\subsection{Formulation}

\begin{figure}[htpb]
\centering
    \includegraphics[scale = 0.8]{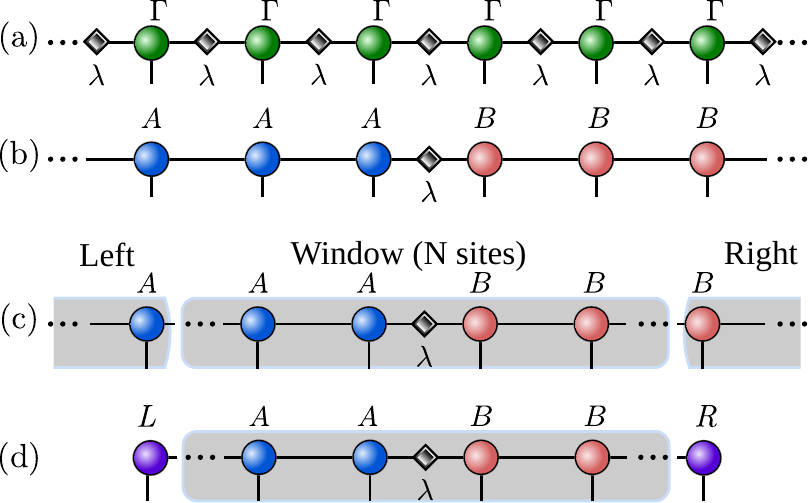}
  \caption{(Color online) (a) Single-site translationally invariant canonical form
    iMPS. (b) One-site translationally invariant mixed canonical form
    iMPS. (c) Partition the whole chain into three parts: the left and
    right semi-infinite sublattices, and the middle part which is
    a window that contains $N$ sites. (d) Finite-size MPS effectively represent the
    iMPS with left- and right-effective sites representing the left
    and right semi-infinite sublattices.}
  \label{fig:figure1}
\end{figure}

Let us consider an infinite-size spin chain for which the wave function is
described by a one-site translationally invariant canonical iMPS
\bea 
\ket{\Psi} =
\sum_{\{s_{i}\}}\ldots\lambda\Gamma^{s_{i-1}}\lambda\Gamma^{s_{i}}\lambda\Gamma^{s_{i+1}}\lambda\Gamma^{s_{i+2}}\ldots\ket{\bf{s}},
\label{eq1}
\eea 
where $\ket{\bf{s}}=\ket{\ldots s_{i-1},s_{i},s_{i+1},s_{i+2}\ldots}$; $s_{i}$ is the local index that represents an element in local
Hilbert space at the $i$th site of the spin chain. The matrices
$\Gamma^{s}$ and $\lambda$ have dimension $\chi\times\chi$ and
$\lambda$ is diagonal. Notice that while the notation of bond dimension is
usually used as $m$ or $D$ in the DMRG language, here we use $\chi$
instead. $\chi$ plays a role as the refinement parameter of the iMPS.
Specifically, the larger the $\chi$ the better the iMPS can represent the
state. Diagrammatically, the iMPS is illustrated in
Fig.~\ref{fig:figure1}(a) where a pair of tensors $\{\Gamma,\lambda\}$ is
repeated at every lattice site throughout the whole infinite chain.

For convenience and later use, we can also rewrite Eq.~\ref{eq1}
in the mixed canonical representation [see Fig.~\ref{fig:figure1}(b)] as
\bea
\ket{\Psi} = \sum_{s}\ldots A^{s_{i-1}}A^{s_{i}}\lambda B^{s_{i+1}}B^{s_{i+2}}\ldots\ket{\bf{s}},
\label{eq2}
\eea
where $A = \lambda\Gamma$ and $B=\Gamma\lambda$ satisfy the left
and right canonical form constraints as follows,
\bea
\sum_{s_{i}}{A^{s_{i}}}^{\dagger}A^{s_{i}}&=&\sum_{s_{i}}{\Gamma^{s_{i}}}^{\dagger}\rho^{R}\Gamma^{s_{i}}=\mathbb{I},
\label{eq3}\\
\sum_{s_{i}}B^{s_{i}}{B^{s_{i}}}^{\dagger}&=&\sum_{s_{i}}\Gamma^{s_{i}}\rho^{L}{\Gamma^{s_{i}}}^{\dagger}=\mathbb{I}.
\label{eq4}
\eea 
In the above equations, $\rho^{R}$ and $\rho^{L}$ are nothing but
the right and left reduced density matrices of the spin chain and
defined as \bea
\rho^{L} &=& \sum_{\al = 1}^{\chi}\big(\lambda_{\al}\big)^{2}\ket{\Phi_{\al}^{L}}\otimes\bra{\Phi_{\al}^{L}},\\
\rho^{R} &=& \sum_{\al =
  1}^{\chi}\big(\lambda_{\al}\big)^{2}\ket{\Phi_{\al}^{R}}\otimes\bra{\Phi_{\al}^{R}},
\eea where $\ket{\Phi_{\al}^{L}}$ and $\ket{\Phi_{\al}^{R}}$ are the
left and right Schmidt vectors that are orthonormal.

The advantage of representing the MPS in the canonical form is that it not only fixes the gauge freedom in the MPS representation,
which would otherwise cause numerical difficulties,
but it is also very convenient for
simplifying the computation of observables of an infinite system. In addition, the
canonical form representation of iMPS is necessary in the truncation
step of time evolution algorithms (both imaginary- and real-time
evolution).

Now, let us partition the whole infinite-size spin chain into three
parts as illustrated in Fig.~\ref{fig:figure1}(c). The middle part, called
the window, contains $N$ sites of the spin chain and the two other
parts contain left and right semi-infinite spin chains attached to
this window.  Then instead of considering a large number of tensors
outside of the window of the iMPS we only use two matrices $L$ and $R$
that represent the whole left and right semi-infinite chains, where
each has dimension $\chi\times\chi$, Fig.~\ref{fig:figure1}(d).
These two matrices represent two boundary sites attached to the window,
and are defined as the infinite boundaries of the finite spin chain.

We have already introduced the idea of shrinking the infinite spin chain to
a finite spin chain with infinite boundary conditions. These infinite
boundary conditions will have to capture all the properties of the
infinite system. Although the idea of shrinking the infinite spin
chain is quite simple, it is more complicated to realize. Specifically,
we need to be sure that our finite-size system with infinite boundary
conditions will behave similarly to the initial infinite system. To
achieve this we require the effective Hamiltonian representing the
infinite system, written in the basis of the finite MPS.

\subsection{Effective Hamiltonian}
Suppose that the total Hamiltonian of the initial infinite spin chain
can be decomposed into five components, written as \bea H = H_{L}+
H_{LW} +H_{W}+H_{WR}+H_{R},
\label{eq5}
\eea where $H_{L}$ and $H_{R}$ are the Hamiltonian components for the
left and right semi-infinite spin chain, $H_{LW}$ ($H_{WR}$) is the
interaction term at the left (right) boundary of the window,
respectively, and finally $H_{W}$ is the Hamiltonian for the window
with $N$ sites.

As we do not consider the whole infinite spin chain, we do not need the
full information contained in the Hamiltonian. Instead, we introduce
the infinite boundary conditions to shrink the infinite chain to the
finite chain. The Hamiltonian for this finite chain will be
effectively described in the same way, as follows: 
\bea \tilde{H} =
\tilde{H}_{L}+\tilde{H}_{LW} +H_{W}+\tilde{H}_{WR}+\tilde{H}_{R},
\label{eq6}
\eea 
where the tilde symbol is added in order to distinguish between
the effective Hamiltonian and the full Hamiltonian of the system. We
can see that $H_{W}$ is the same in both Eq.~\ref{eq5} and
Eq.~\ref{eq6}. Our task is to find the effective Hamiltonians of the
left and right semi-infinite chain and their interaction components
with the window [the components in Eq.~\ref{eq6} with the tilde symbol].

We now show the method to calculate the effective Hamiltonian by using
spin-1 isotropic antiferromagnetic Heisenberg model as an example. The
Hamiltonian contains nearest-neighbor interaction terms as follows:
\bea H= \sum_{i}\vec{S}_{i} \cdot \vec{S}_{i+1},
\label{eq7}
\eea where $\vec{S} = (S^{x},S^{y},S^{z})$ is the vector containing
matrices for the spin-1 representation of the spin algebra.

The effective Hamiltonian can now be written as
\bea
\tilde{H}&=&\tilde{H}_{L}+\tilde{\vec{S}}_{L} \cdot \vec{S_{1}}
+\sum_{i=1}^{N-1}\vec{S}_{i} \cdot \vec{S}_{i+1}
+\vec{S}_{N} \cdot \tilde{\vec{S}}_{R}
+\tilde{H}_{R}.
\eea 
We need to find the left and right effective Hamiltonians
$\tilde{H}_{L}, \tilde{H}_{R}$ and also operators
$\tilde{\vec{S}}_{L},\tilde{\vec{S}}_{R}$ which are
$\chi\times\chi$ matrices. The procedure to obtain the effective
Hamiltonian is described in detail in Ref. \onlinecite{Ian1}, and we
now briefly review it here.

\begin{figure}[htpb]
  \centering
    \includegraphics[scale = 0.8]{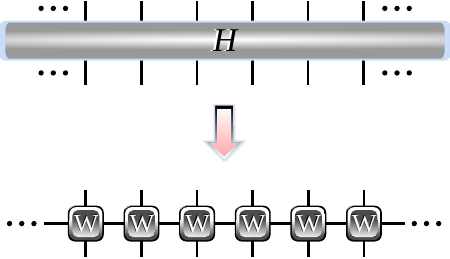}
  \caption{(Color online) The full Hamiltonian of the system is decomposed into the tensor
    product of local matrix product operators.}
  \label{fig:figure2}
\end{figure}

Let us introduce the infinite matrix product operator (iMPO) which
has the following form for an infinite-size spin chain.  
\bea
\bra{\sigma}H\ket{\sigma'} =\ldots W^{s_{i}s'_{i}}W^{s_{i+1}s'_{i+1}}\ldots, 
\eea
where we denote $\ket{\sigma} = \ket{\ldots s_i, s_{i+1} \ldots}$ as the basis of
the system. As the unit cell of this model contains a single site, the iMPO is
represented by the same matrices $W^{ss'}$ repeated at every site
of the chain; see Fig.~\ref{fig:figure2}.

With each type of Hamiltonian there are several ways to construct the
iMPO; here we are using the method proposed in Ref. \onlinecite{Ian}
where all the matrices are in lower triangular forms. For the
Hamiltonian described by Eq.~(\ref{eq7}), these matrices have the
following form:
\[
W = \begin{bmatrix}
  \mathbb{I} & 0 & 0 & 0 & 0\\
  S^{x} & 0 & 0 & 0 & 0\\
  S^{y} & 0 & 0 & 0 & 0\\
  S^{z} & 0 & 0 & 0 & 0\\
  0 & S^{x} & S^{y} & S^{z} & \mathbb{I}\\
\end{bmatrix},
\]
where $\mathbb{I}$ is a $3\times3$ identity matrix.

\begin{figure}[htpb]
  \centering
  \includegraphics[scale = 0.8]{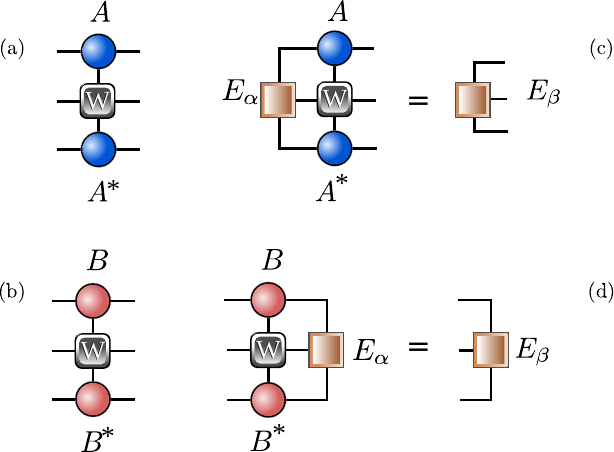}
  \caption{(Color online) (a) Generalized transfer matrix for finding the left dominant
    eigenvector $\vec{E}$. (b) Generalized transfer matrix for finding the right dominant
    eigenvector $\vec{E}$. (c) Equation to find left dominant eigenvector.
    (d) Equation to find right dominant eigenvector.}
  \label{fig:figure3}
\end{figure}

We now review the scheme proposed in Ref. \onlinecite{Ian1} to find
all the left effective operators; a similar scheme can be applied for
the right operators. Specifically, we need to find the dominant eigenvector
of the transfer matrix diagrammatically illustrated in
Fig.~\ref{fig:figure3}(a). This dominant eigenvector contains five
components, $\vec{E} = (E_{1}, E_{2}, E_{3}, E_{4}, E_{5})$. As we
will see later, this dominant eigenvector contains the information of
the left effective Hamiltonian that we need, or in DMRG terminology,
$\vec{E}$ is the vector of block operators describing the effective Hamiltonian.
However, as the transfer matrix is not diagonalizable, we need to find
all the elements of $\vec{E}$ independently by employing the recursion
relation, see Fig.~\ref{fig:figure3}(c), which reads \bea E_{\al}(n+1)
=T_{W_{\al\al}}(E_{\al}(n))+\sum_{\bet>\al}T_{W_{\bet\al}}(E_{\bet}(n)),
\label{eq8}
\eea where we have defined 
\bea T_{X}(E)
=\sum_{ss'}{\braket{s}{X|s'}A^{s'\dagger}EA^{s}},
\eea
which is the generalized transfer operator to include a local operator
$X$ acting on the physical degree of freedom of the MPS. The relevant local
operators will be obtained from the elements of the MPO matrix
$W_{\beta\alpha}$, and we make use of the fact that $W$ is lower triangular
to restrict the summation to $\beta \geq \alpha$.
Since the other terms $S_{\beta\alpha}$ with $\beta<\alpha$
are equal to zero, we can
solve immediately the recursion relation Eq.~\ref{eq8} for the last component, 
in this example being $E_5$, 
\bea
E_{5}(n+1) &=& T_{W_{55}}(E_{5}(n))\nn\\
& =&\sum_{ss'}{\braket{s}{\mathbb{I}|s'}A^{s'\dagger}E_{5}(n)A^{s}},
\eea 
which implies that in the large-$n$ limit $E_5(n)$ is the eigenvector of the transfer operator
with largest eigenvalue.
If the iMPS is in the canonical form then this largest eigenvalue will be 1 and
we have \bea
E_{5}  = \tilde{\mathbb{I}}, \eea
where $\tilde{\mathbb{I}}$ is a $\chi\times\chi$ identity matrix.
Moving on to $E_4$, we have \bea
E_{4}(n+1) &=& \underbrace{T_{W_{44}}(E_{4}(n))}_{0}+\sum_{ss'}{\braket{s}{S^{z}|s'}A^{s'\dagger}\underbrace{E_{5}(n)}_{\tilde{\mathbb{I}}}A^{s}}\nn\\
&=&
\sum_{ss'}{\braket{s}{S^{z}|s'}A^{s'\dagger}A^{s}}=\tilde{S}_{L}^{z},
\eea 
and here the fact that the diagonal matrix element $W_{44} = 0$ implies that
the solution for $E_4$ is simply a function of $E_5$ and local operators.  Similarly,
\bea
E_{3}(n+1) &=& \underbrace{T_{W_{33}}(E_{3}(n))}_{0}+\sum_{ss'}{\braket{s}{S^{y}|s'}A^{s'\dagger}\underbrace{E_{5}(n)}_{\tilde{\mathbb{I}}}A^{s}}\nn\\
&=&\sum_{ss'}{\braket{s}{S^{y}|s'}A^{s'\dagger}A^{s}}=\tilde{S}_{L}^{y},
\eea \bea
E_{2}(n+1) &=& \underbrace{T_{W_{22}}(E_{2}(n))}_{0}+\sum_{ss'}{\braket{s}{S^{x}|s'}A^{s'\dagger}\underbrace{E_{5}(n)}_{\tilde{\mathbb{I}}}A^{s}}\nn\\
&=&\sum_{ss'}{\braket{s}{S^{x}|s'}A^{s'\dagger}A^{s}}=\tilde{S}_{L}^{x},
\eea and finally, the most complicated term that contains the
effective Hamiltonian of the left semi-infinite spin chain is
determined as \bea
E_{1}(n+1) &=& T_{W_{11}}(E_{1}(n))+\sum_{\bet>1}T_{W_{\bet 1}}(E_{\bet}(n))\nn\\
&=& \sum_{ss'}{\braket{s}{\mathbb{I}|s'}A^{s'\dagger}E_{1}(n)A^{s}}\nn\\
&&+\sum_{ss'}{\braket{s}{S^{x}|s'}A^{s'\dagger}\underbrace{E_{2}(n)}_{\tilde{S}_{L}^{x}}A^{s}}\nn\\
&&+\sum_{ss'}{\braket{s}{S^{y}|s'}A^{s'\dagger}\underbrace{E_{3}(n)}_{\tilde{S}_{L}^{y}}A^{s}}\nn\\
&&+\sum_{ss'}{\braket{s}{S^{z}|s'}A^{s'\dagger}\underbrace{E_{4}(n)}_{\tilde{S}_{L}^{z}}A^{s}}.
\label{eq9}
\eea We can also write this equation in a compact form as \bea
E_{1}(n+1)&=&\sum_{ss'}{\braket{s}{\mathbb{I}|s'}A^{s'\dagger}E_{1}(n)A^{s}}+C,
\label{eq10}
\eea where $C$ is a constant that is defined as the summation of last
three terms in Eq.~\ref{eq9}.
Our task is to solve Eq.~\ref{eq10}. To see how this is done, let
us assume the initial solution $E_{1}(0) = 0$.  This is an arbitrary choice that
has no effect on the final solution, up to an irrelevant constant. Then, \bea
E_{1}(1) &=& C\nn\\
E_{1}(2) &=& T_{\mathbb{I}}(C)+C\nn\\
E_{1}(3) &=& T_{\mathbb{I}}(T_{\mathbb{I}}(C)+C)+C \\
         &=& T_{\mathbb{I}}(T_{\mathbb{I}}(C))+T_{\mathbb{I}}(C)+C\nn\\
&\ldots&\nn\\
E_{1}(n+1) &=& T_{\mathbb{I}}(E_{1}(n))+C
\label{recursion}
\eea In general we can write the solution as follows,
\bea
E_{1}(n) &=&\sum_{k = 0}^{n-1} T^{k}_{\mathbb{I}}(C)\nn\\
&=& C + T_{\mathbb{I}}(C) + T_{\mathbb{I}}(T_{\mathbb{I}}(C)) 
+ T_{\mathbb{I}}(T_{\mathbb{I}}(T_{\mathbb{I}}(C)))+\ldots\nn\\
\eea 
This is the summation of a geometric series, which has the solution
\bea 
\sum_{k = 0}^{n-1}ax^{k} &=&\frac{a(1-x^{n})}{1-x}.  
\eea
In our case 
\bea \sum_{k = 0}^{n-1} T^{k}_{\mathbb{I}}(C)
&=&\frac{(\tilde{\mathbb{I}}-T_{\mathbb{I}}^{n})(C)}{(\tilde{\mathbb{I}}-T)(C)}.
\eea Notice that the spectrum of transfer matrix $T_{\mathbb{I}}$ will
contain the identity $\tilde{\mathbb{I}}$ and density matrix $\tilde{\mathbb{\rho}}$
as a left/right eigenvector pair with eigenvalue 1. 
Therefore, this summation
will be diverging. To avoid this, let us decompose the summation into two terms
as \bea E_{1}(n) = \tilde{H}_L + e_{0}n\tilde{\mathbb{I}},
\label{eq11}
\eea where $\tilde{H}_L$ contains all the terms that are perpendicular to
the identity (meaning $\tr \tilde{H}_L \rho = 0$)
and is actually the effective Hamiltonian of the left
semi-infinite chain; $e_{0}$ is a constant equal to the energy
per site of the infinite chain. Note that $\tilde{H}_L$ removes the
constant contribution of the energy that would diverge in the thermodynamic limit.
Now we can check the recursion
relation by substituting Eq.~\ref{eq11} into Eq.~\ref{recursion};
we have 
\bea \tilde{H}_L + e_{0}(n+1)\tilde{\mathbb{I}}
&=&T_{\mathbb{I}}(\tilde{H}_L)+T_{\mathbb{I}}(e_{0}n\tilde{\mathbb{I}})+C
\eea 
This simplifies to a linear equation for $\tilde{H}_L$,
\bea
(\tilde{\mathbb{I}}-T_{\mathbb{I}})(\tilde{H}_L)&=&C-e_{0}\tilde{\mathbb{I}}.
\label{eq12}
\eea
where $e_{0} = Tr(\rho C)$ ($\rho$ is density matrix). By solving this linear 
equation we find the effective Hamiltonian $\tilde{H}_L$ and this completes the
vector of block operators $E_\alpha(n)$. Note that
the energy per site contribution $e_0 n \tilde{\mathbb{I}}$ is a
constant shift in the energy and is therefore irrelevant for most purposes.

In summary, we have explained in this section how to obtain the
effective Hamiltonian on the left of the window. Specifically, we
have obtained $\tilde{H}_{L} = E_{1}$ and also operators
$\tilde{\Vec{S}}_{L}=\{\tilde{S}_{L}^{x}, \tilde{S}_{L}^{y},
\tilde{S}_{L}^{z}\} = \{E_{2}, E_{3}, E_{4}\}$. For the right
effective Hamiltonian, a completely similar procedure is
performed. In the next section we will use this calculation to
investigate the problem of real-time evolution of iMPS in the presence
of local perturbation.

\section{Application: Real-time evolution of iMPS in the presence of local perturbation}
We now apply the procedure for finding the effective
Hamiltonian and the infinite boundary conditions proposed above to
study dynamical properties of an infinite spin chain in the presence
of a local perturbation. As an infinite MPS will be effectively represented by a
finite MPS, we can apply a standard MPS time-evolution technique to
study the reaction of the infinite system to a local perturbation. The MPS
technique that we use here is the TEBD algorithm.

\subsection{Local perturbation}
\begin{figure}[h!]
  \centering
  \includegraphics[scale = 0.5]{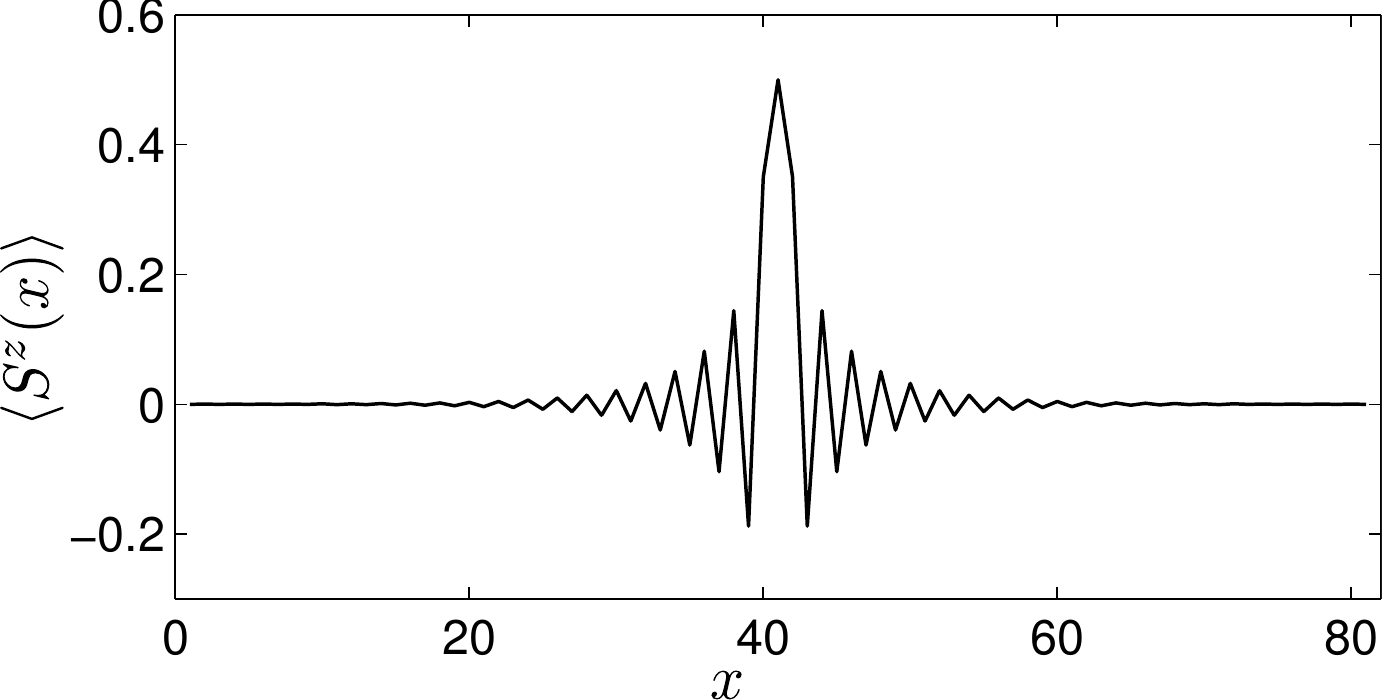}
  \caption{The wave packet (local magnetization) is formed after
    flipping one spin in the middle of the chain. The result is
    obtained by using a two-site translationally invariant iMPS for
    the ground state with bond dimension $\chi = 160$.}
  \label{fig:figure4}
\end{figure}

We wish to take an infinite spin chain which is in its ground state,
and perturb locally one site. Suppose that we have
already found the ground state of the system (for example,
by iDMRG or iTEBD), represented by a
translationally invariant iMPS with a one- or two-site unit cell; the wave function
$\ket{\Psi_{GS}}$ is written as in Eq.~\ref{eq2}. Then we choose one
site and perturb it locally by flipping the spin of that site with
flipping spin operators $S^{+}$ (flip spin up) or $S^{-}$ (flip spin
down). The system is not in the ground state anymore, but a mixture of
excited states, and is no longer described by a translationally invariant
iMPS. Let us flip the spin at a certain position $j$ in the chain and
define a new state as 
\bea 
\ket{\tilde{\Psi}}&=& S^{+}_{j}\ket{\Psi_{GS}}.  
\eea 
As a result of spin flipping, a wave
packet is formed centered at the flipped spin. As an
illustration using the spin-1 isotropic antiferromagnetic Heisenberg
model, in Fig.~\ref{fig:figure4}, we plot the local magnetization
of the system after flipping one spin in the middle of the chain. We
can see that a wave packet is formed with the peak located in the
middle site. The amplitude of this wave packet decreases when moving
away from the middle point. The width of the wave packet depends on
the correlation length of the system. Note however that despite
the breaking of translational invariance at long range, only
one tensor of the MPS is different from that of the 
translationally invariant ground state.

\subsection{Real-time evolution}

Let us now study real-time evolution of an infinite spin chain. The
initial state of the system is a locally perturbed state
$\ket{\tilde{\Psi}(0)}$. This state will evolve in time and is
described by the solution of the Schroedinger equation \bea
\ket{\tilde{\Psi}(t)} = e^{-i\tilde{H}t}\ket{\tilde{\Psi}(0)}.  \eea
As mentioned above, this state can be effectively represented by a 
finite MPS
containing $N+2$ sites where the perturbed site is in the middle of the
chain at site $i = N/2$, with an effective Hamiltonian $\tilde{H}$ describing the finite
system. The two boundary sites are now represented by the boundary tensors $L^\alpha$ and
$R^\beta$, which are the usual boundary sites
of a finite MPS with dimensions $1 \times \chi$ and $\chi \times 1$ respectively, 
except now the local Hilbert space is the $\chi$-dimensional effective Hilbert
space for the left and right semi-infinite strips.
In practice we do not actually need the $L^\alpha$ and $R^\beta$ tensors as these
are identity elements; $L^\alpha_i = \delta_{\alpha i}$ and $R^\beta_j = \delta_{\beta j}$,
but their use allows us to formally write the state of the system 
Eq.~\ref{eq2} as a finite MPS,
\bea
\ket{\tilde{\Psi}} =
\sum_{\{s_{i}\}} L^{\alpha} A_1^{s_{1}} \lambda A_2^{s_{2}}\ldots
A_N^{s_{N}} R^{\beta} \: \ket{\alpha, \mathbf{\tilde{s}}, \beta},
\eea
where $\ket{\bf{\tilde{s}}}=\ket{s_{1}, s_{2}, \ldots s_{N}}$.,
The location of the $\lambda$ matrix will sweep through the system
as usual in finite-size DMRG algorithms,\cite{DMRGMPS} with
all the tensors to the left of $\lambda$ satisfying the
the left canonical constraint of Eq.~\ref{eq3} and all
the tensors on the right of $\lambda$ matrix satisfying the right
canonical constraint in Eq.~\ref{eq4}. Note that it is not possible
to write this system in the canonical $\Gamma,\Lambda$ form used by Vidal\cite{Vidal1}
without modifying the boundary tensors $L,R$.

With an effective finite system representing the infinite system,
we can proceed with the real-time evolution by employing the TEBD
algorithm. Before continuing, we will briefly reiterate the main
features of the TEBD algorithm. For more details, refer to
the original work\cite{Vidal1}. In this algorithm the time evolution
operator $e^{-iHt}$ is decomposed as a product of $M$ operators
$e^{-iH\delta t}$ (where $\delta t \ll 1$ is the small time step and
$M = t/\delta t$). In turn, each term $e^{-iH\delta t}$ is decomposed
into products of local terms by using Suzuki-Trotter decomposition.
Normally, the Hamiltonian is written as the summation of two terms.
With the Hamiltonian just containing nearest-neighbor
interaction terms, we can rewrite it in the following form:
\bea H=H_{odd}+H_{even}, 
\eea
 where $H_{odd} =\sum_{odd~i}h^{[i,i+1]}$ and
$H_{even} =\sum_{even~i}h^{[i,i+1]}$. Terms in either
$H_{odd}$ or $H_{even}$ commute with each other. However, the terms in
$H_{odd}$ do not commute with ones in $H_{even}$ in general. Then the
first-order Suzuki-Trotter decomposition of the time evolution
operator at each time step $\delta t$ is 
\bea
e^{-iH\delta t} &=& e^{-iH_{odd}\delta t}e^{-iH_{even}\delta t} + O(\delta t^2)\nn\\
&=&\bigotimes_{odd~i} e^{-ih^{[i,i+1]}\delta  t}\bigotimes_{even~i}e^{-ih^{[i,i+1]}\delta t} + O(\delta t^2).\nn\\
\eea 
As a consequence of the non zero commutation relation between the odd
and even terms of the Hamiltonian, the Suzuki-Trotter decomposition
will produce some error on the order of $\delta t^2$. However,
this error can be controlled by using a small time step $\delta t$ or by
taking high-order decomposition.\cite{Suzu1}

Here, we modify slightly the TEBD algorithm to investigate the
real-time evolution of our locally perturbed system.  Specifically, we
do not need to find the inverse of the $\lambda$ matrix after acting the
two-body gate on a given link of two sites, but instead we use two more SVDs to shift the $\lambda$ matrix by two sites to
the next update link. This step is also important to get an 
optimal truncation which is essential for each local update. 
As a price of implementing two extra SVDs, this step may be
a little bit costly. However, the big advantage of doing this is that we
can avoid the inverse of $\lambda$ that is numerically unstable.

For convenience and clarity, we write the time evolution operator with time step $\delta t$ in the first order
of Suzuki-Trotter decomposition:
\bea
e^{-i\tilde{H}\delta t} &=& e^{-i(\tilde{H}_{L} + \tilde{H}_{LW} + H_{W} + \tilde{H}_{R} +\tilde{H}_{RW})\delta t}\nn\\
&\cong&e^{-i\tilde{H}_{L}\delta t}e^{-i \tilde{H}_{LW}\delta t}e^{-iH_{W}\delta t}e^{-i \tilde{H}_{R}\delta t}e^{-i\tilde{H}_{RW}\delta t}\nn\\
&=&U_{L}U_{LW}U_{W}U_{R}U_{RW}.  
\eea
\begin{figure}[htpb]
  \centering
  \includegraphics[scale=0.72]{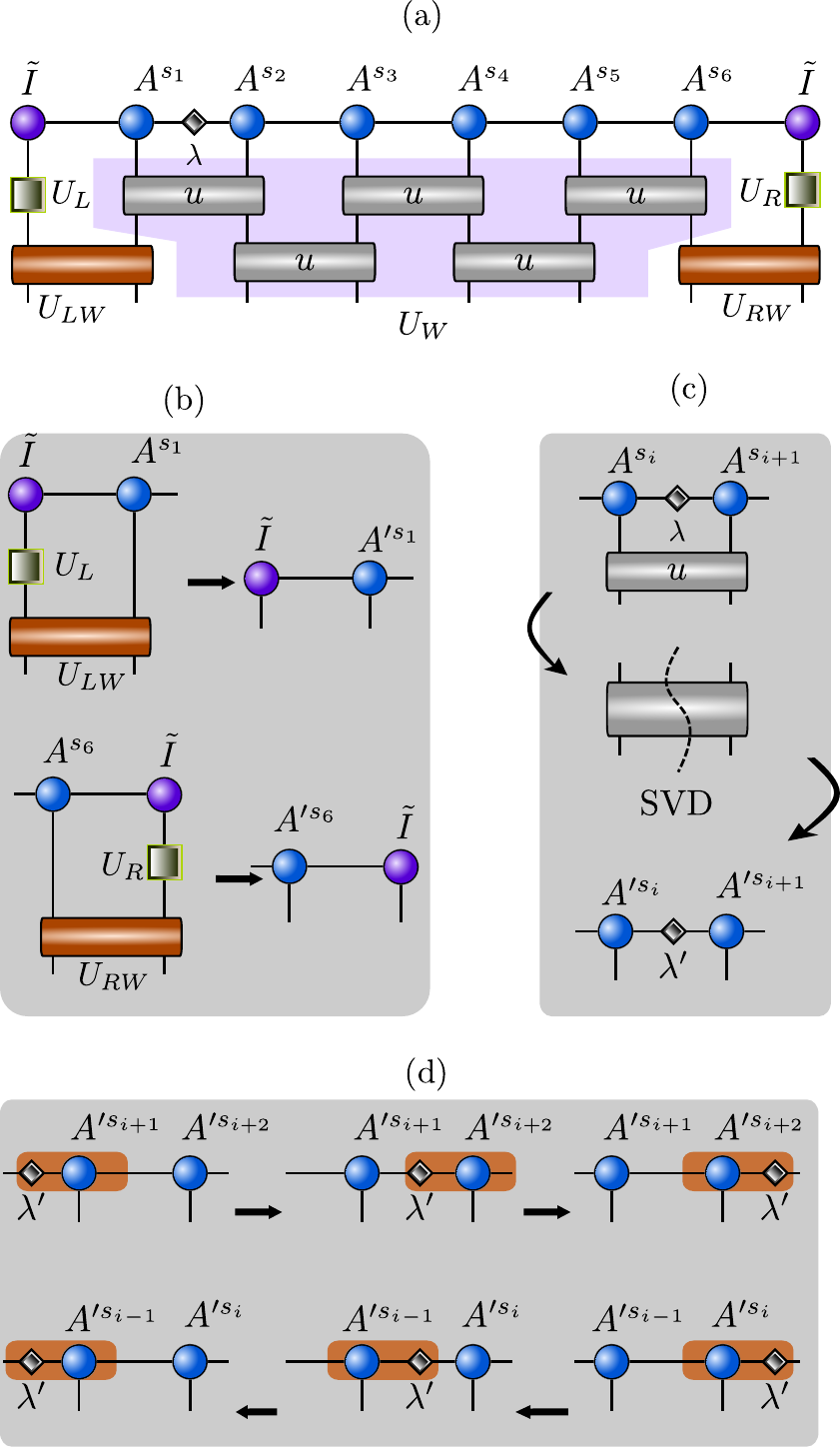}
  \caption{(Color online): (a). Applying the operator $e^{-i\tilde{H}\delta t}$ to
    the effective finite MPS. (b). Update the left and right tensors
    by contracting tensors $\{U_{L}, U_{LW}, \tilde{I}, A^{s_{1}}\}$
    and $\{U_{R}, U_{RW}, \tilde{I}, A^{s_{6}}\}$. (c). Update the new
    tensors when applying two-body gate $u$ on the odd or even link.
    Contracting all the tensors involved and take the SVD of that. The
    bond dimension will increase after taking the SVD, so we need to do
    the truncation to keep new tensors $A'^{s_{i}}, A'^{s_{i+1}}$, and
    $\lambda'$ in the desired bond dimension. (d). Shifting
    $\lambda'$ to the right (or left) of the updated link if sweeping
    direction is from left to right (or right to left) by taking two
    successive SVDs.}
  \label{fig:figure5}
\end{figure}
The update scheme of real-time evolution at each time step is
illustrated diagrammatically in Fig.~\ref{fig:figure5} for an $N=6$
window. However, in principle $N$ can be any arbitrary finite number.
Each time step includes two successive sweeps: one from the left to
the right and vice versa. Note that with our specific choice of an even
number of sites inside the window $N$ the interaction terms on the
left and right sides of the window are operationally equivalent to the
even terms.
\section{Results}
By replacing the iMPS with an effective finite MPS containing two
boundary sites we can evolve the locally perturbed ground state in
time. Here we present the results computed for the spin-1 isotropic
antiferromagnetic Heisenberg model. The initial ground state
is represented by the iMPS with bond dimension $\chi = 160$ and is
evolved to a state with maximum truncated bond dimension $\chi_{C} =
200$. Truncation error is approximately equal to $10^{-7}$. After
flipping the central spin, the system is evolved up to time $t = 30$ where
time step $\delta t = 0.05$ is used for the fourth-order
Suzuki-Trotter decomposition.
\subsection{Wave packet propagating in time}
\begin{figure}[htpb]
  \centering
  \includegraphics[scale = 0.7]{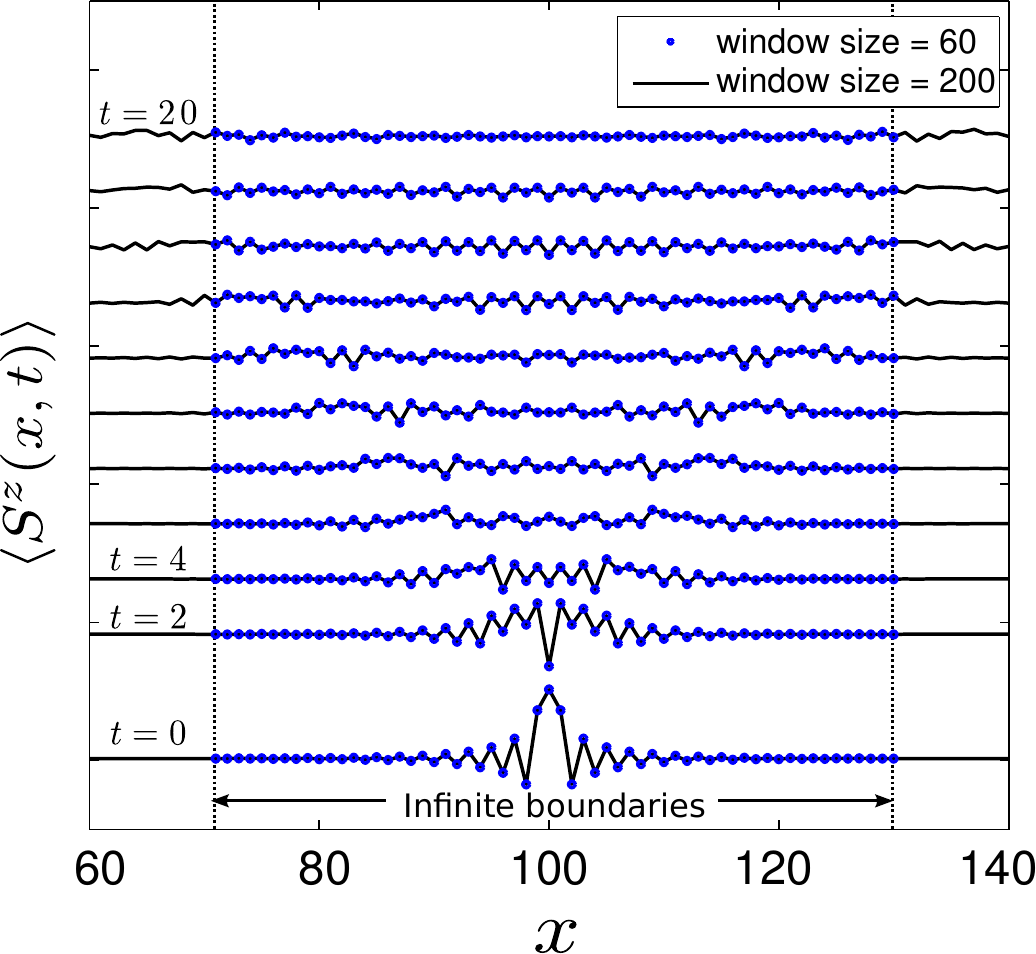}
  \caption{(Color online) Wave packet propagates in time with window size $N = 60$
    (blue dots). The window is expanded to $N=200$ (black solid lines)
    after simulation to see how the wave front propagates beyond the
    infinite boundaries.}
  \label{fig:figure6}
\end{figure}

In order to understand how the wave packet is propagating in the
effective finite MPS where the infinite boundaries are present, we
compute the local magnetizations in time at each site of the spin
chain. The result is shown in the Fig.~\ref{fig:figure6}.
For the system with window size $N=60$, as we can see, the wave
packet at the beginning is formed at the middle of the chain and then
spreads out. Importantly, what we can see here is that when the wave front
hits the infinite boundaries, there is no back reflection or
counter propagating effect; it passes through the boundaries.
This can be verified when we look at how the wave packet propagates
outside the window in the effective Hilbert space. Specifically, we
expand the window after the simulation by inserting the original
orthogonal tensors $A$ and $B$ at the edges. From this we calculate
expectation values outside the original window. From
Fig.~\ref{fig:figure6} we can see that when the window is
expanded to window size $N = 200$, the wave front moves smoothly to the
exterior, justifying our approach.

\begin{figure}[htpb]
  \centering
  \includegraphics[scale = 0.5]{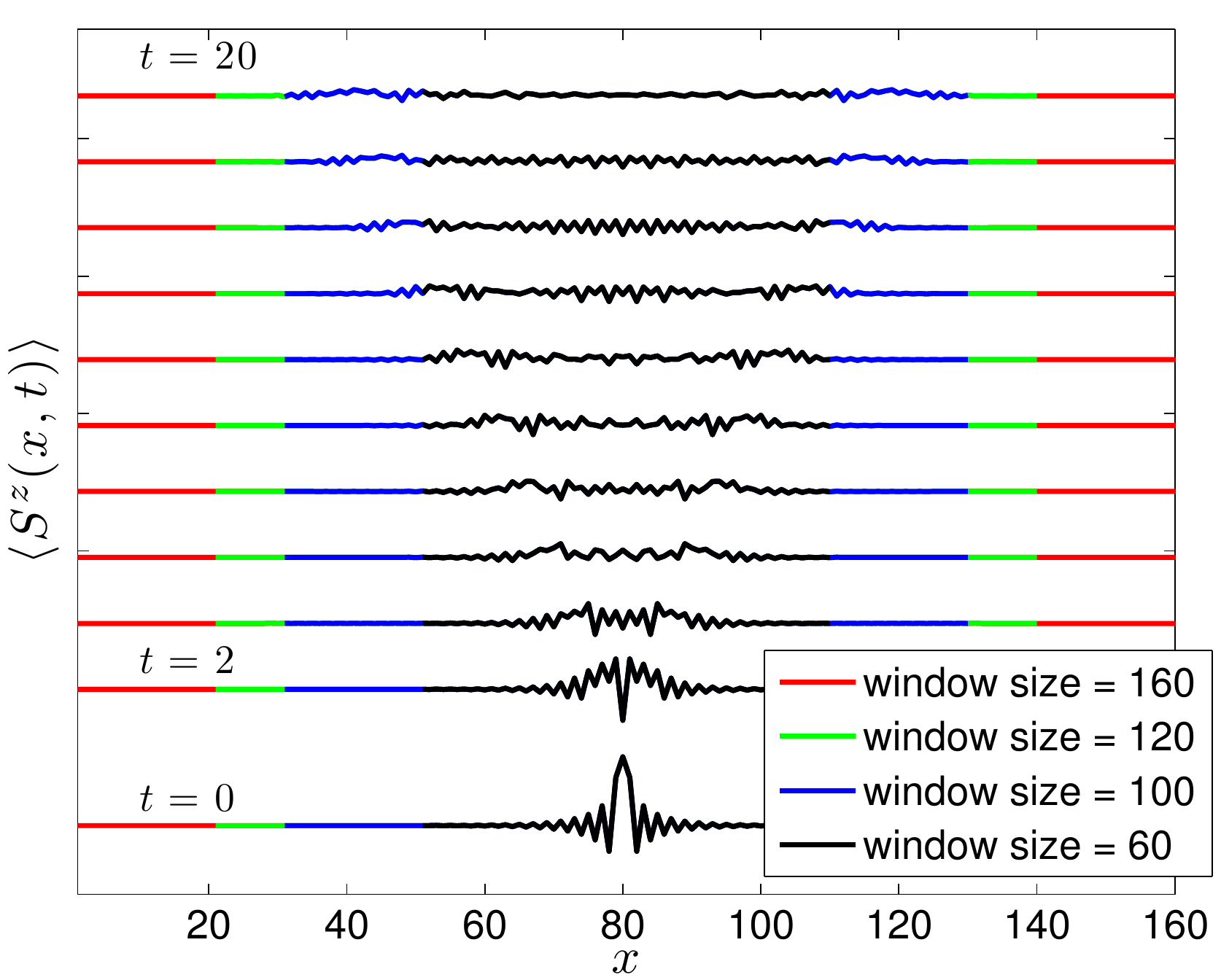}
  \caption{(Color online) Wave packet propagates in time with different window sizes
    which are fixed at the beginning of real-time evolution. The lines
    are distinguishable in the center of the plot.}
  \label{fig:figure7}
\end{figure}

\begin{figure}[htpb]
  \centering
  \includegraphics[scale = 0.4]{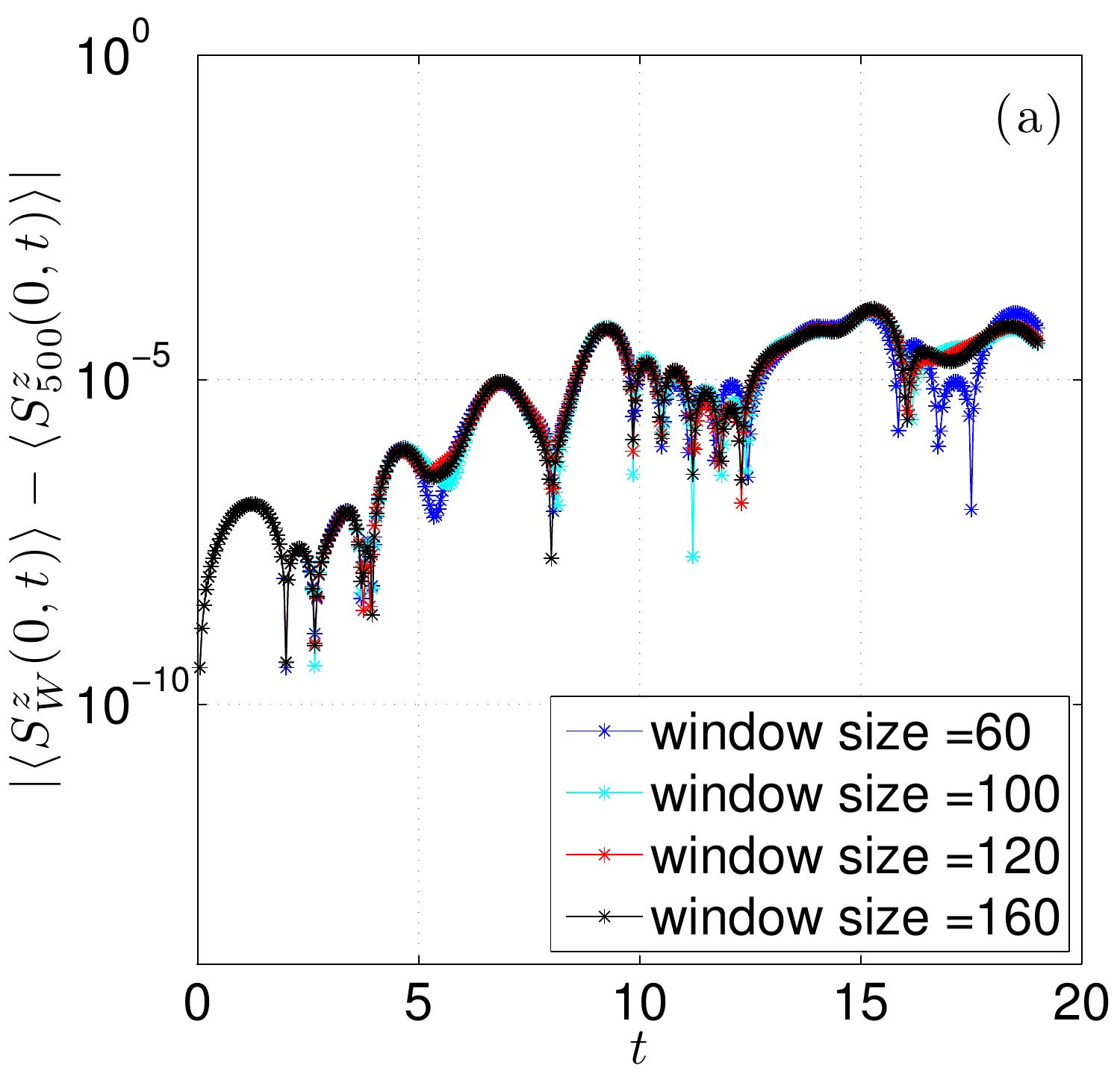}\\
  \includegraphics[scale = 0.4]{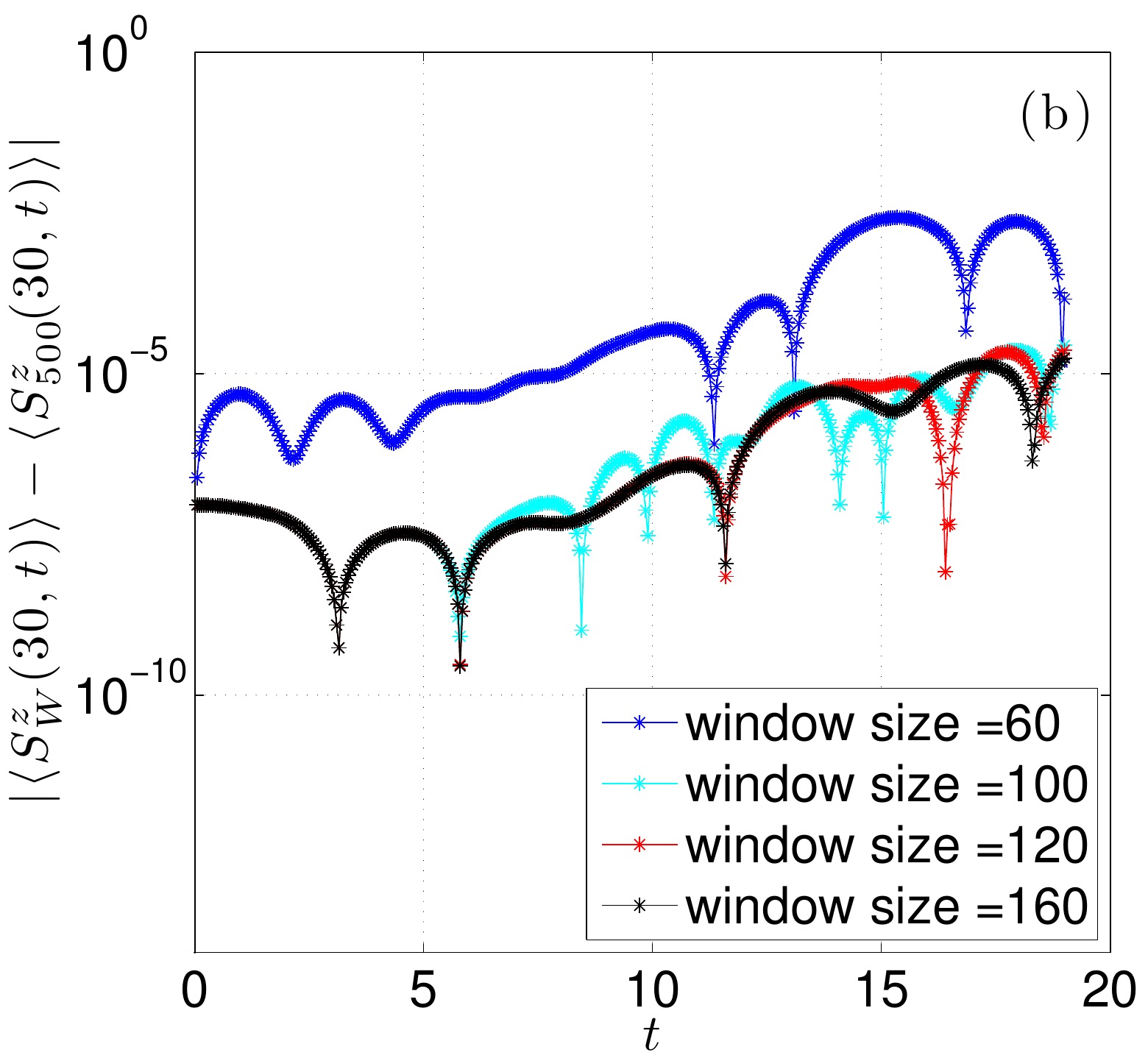}
  \caption{(Color online): Comparisons of the differences of local magnetization in
    time at a fixed site $x_{i}$ between window sizes $N =
    \{60,100,120,160\}$ and a highly accurate calculation using a 500-site finite chain. 
    (a). Perturbed point of chain $x_{i} = 0$. (b) Boundary point of the chain $x_{i} = 30$.}
  \label{fig:figure8}
\end{figure}

Now, for comparison, we also plot the propagation of the wave packet in
time with different window sizes. These windows have sizes fixed from
the beginning of the real-time evolution. As we can see from
Fig.~\ref{fig:figure7}, the wave packets of
different window sizes are coincident with each other in the middle
region of the plot. This is exemplified in Fig.~\ref{fig:figure8}(a) where
we plot the error in the magnetization at the site of the perturbation as a function
of time $t$, for window sizes 60, 100, 120, 160. The curves are nearly coincident,
showing that the dominant contribution to the error is from the Suzuki-Trotter 
decomposition, not the finite window. Figure~\ref{fig:figure8}(b) shows the error in
the magnetization 30 sites away.
If the window is larger than 60 sites, then we again see no error
beyond the usual Suzuki-Trotter error. For the 60-site window,
the site where we measure the magnetization corresponds to the edge of the window
and in this calculation the error is somewhat increased, even at very small times,
which is probably due to a slight mismatch of using a slightly different approximation
for the time evolution operator inside the window (Suzuki-Trotter) and outside the window
(direct calculation of the exponential of the effective Hamiltonian). Nevertheless,
there is no sign of any significant increase in the relative error due to the wave front
passing through the edge of the window and into the infinite boundary tensor. Indeed,
the leading edge of the wave front passes site 30 at around $t=10$, and by $t=18$
the entire wave front has already passed.

\subsection{Unequal-time two-point correlator and spectral function}
Let us define an unequal-time two-point correlator as \bea A(x,t)
=\braket{\phi|S_{x}^{-}(t)S_{x_{M}}^{+}(0)}{\phi}, \eea where the
subscripts in spin-flip operators indicate positions of the chain and
$x_{M}$ is the middle position; $\ket{\phi}$ is the initial state of
the system that we want to evolve. This equation is equivalent with
\bea A(x,t) =e^{iE_{G}t}\braket{\phi|S_{x}^{-}(0)}{\psi(t)},
\label{eq13}
\eea in which we have already replaced $S^{-}_{x}(t) =
e^{iHt}S^{-}_{x}(0)e^{-iHt}$ and $\ket{\psi(t)} =
e^{-iHt}S^{+}(0)\ket{\phi}$. We also have a phase factor appear in
Eq.~\ref{eq13} due to $\ket{\phi}$ being the eigenvector of the
Hamiltonian $H$ corresponding to the eigenvalue $E_{G}$. Obviously,
the unequal-time two-point correlator $A(x,t)$ can be calculated
easily, as the time-evolved state $\ket{\psi(t)}$ can be obtained
quickly from the scheme proposed above for evolving the locally
perturbed state.

\begin{figure}[htpb]
  \centering
  \includegraphics[scale = 0.2]{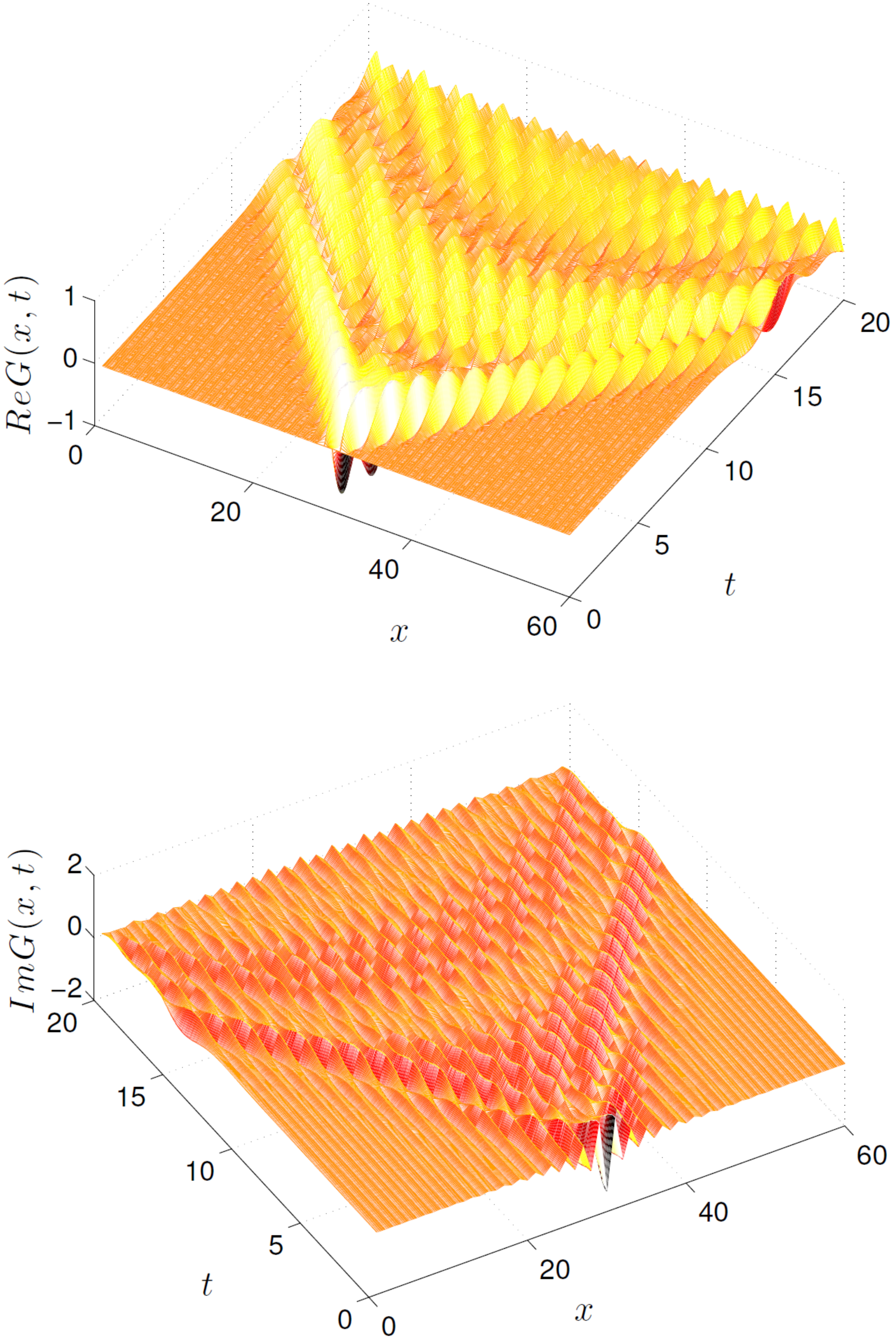}\\
  \caption{(Color online) Plots of the real and imaginary parts of the Green's
    function versus time and spin chain space.}
  \label{fig:figure9}
\end{figure}

From the unequal-time two-point correlator we construct the Green's
function that is defined as \bea G(x,t) = -iA(x,t).  \eea
Figure~\ref{fig:figure9} shows the plots of the real and imaginary parts
of the Green's function for the system with window size $N=60$. As we
can see, there are wave fronts propagating from the middle point
toward the infinite boundaries. Again, there is no back reflection of
the wave front at the boundaries.

By Fourier transforming of the Green's function into momentum and
frequency spaces, we can extract the spectrum of the lattice system.
Specifically, the Fourier transform of $G(x,t)$ is \bea G(q,\omega) =
\int_{-\infty}^{\infty}dt~e^{i\omega t}\sum_{x}e^{-iqx}G(x,t).
\label{eq14}
\eea For the case of the spin-1 isotropic antiferromagnetic Heisenberg
model, the Green's function is even in $x$ and $t$, and we can simplify
Eq.~\ref{eq14} as follows:
\bea G(q,\omega)=\int_{-\infty}^{\infty}dt~\cos{\omega t}\sum_{x}\cos{qx}G(x,t).
\label{eq15}
\eea $G(x,t)$ is a continuous function in time $t$. However, in our
simulation, we have already discretized the time into the small time
steps $\delta t$. Therefore, Eq.~\ref{eq15} can be written as \bea
G(q,\omega)\approx 2\sum_{t=0}^{T_{max}}\cos{\omega
  t}\sum_{x}\cos{qx}G(x,t).
\label{eq16}
\eea The spectral function is now defined as \bea S(q,\omega) =
-\frac{1}{\pi}\text{Im}G(q,\omega).  \eea 
Note that as we have already introduced the infinite boundaries for our finite MPS, the
wave front can now propagate freely through these boundaries without
any back reflection. In principle, in obtaining the spectral function
we do not need to have any cutoff in time to keep the available data.
\begin{figure}[htpb]
  \centering
  \includegraphics[scale = 0.4]{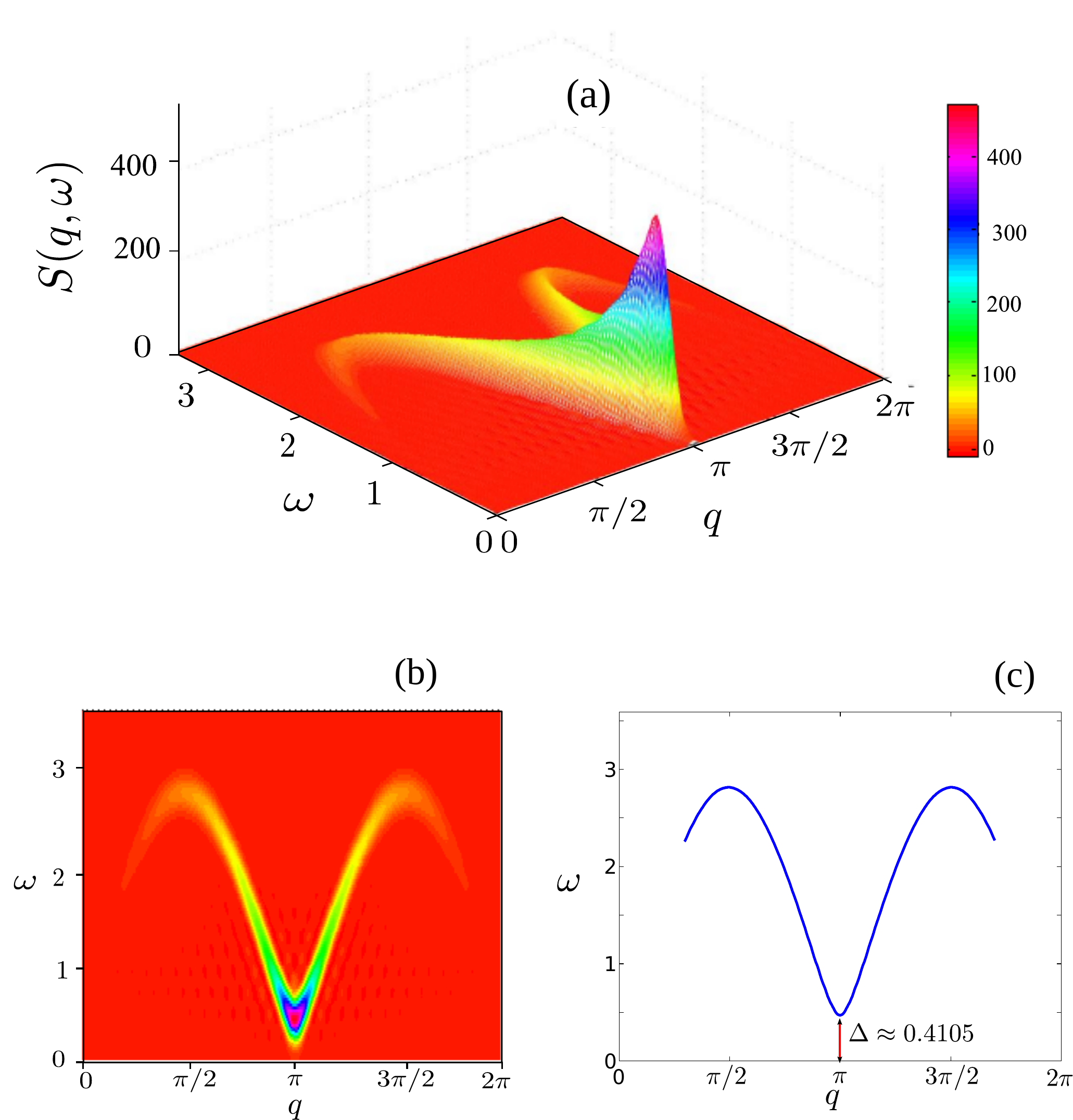}\\
  \caption{(Color online) (a) Spectral function versus momentum and frequency for
    the spin-1 isotropic Heisenberg model; the window is $N = 60$.
    (b) Spectrum viewed from the top when it is projected on the
    $(\omega,q)$ plane. (c) The dispersion relation is derived from
    the maximum of the spectrum.}
  \label{fig:figure10}
\end{figure}
In Fig.~\ref{fig:figure10}(a), we plot the spectral function versus
momentum and frequency. In order to get a smooth spectrum, we have
multiplied $G(x,t)$ with a Gaussian window function of the form $\text{exp}[-4(t/T)^{2}]$ as introduced in
Ref.~\onlinecite{White3}. By viewing from the top of this figure, we
can see the dispersion relation appears very clearly in
Fig.~\ref{fig:figure10}(b). Collecting the data pairs $(\omega,q)$ that correspond to the maximum of spectrum and plotting them, 
we can
see the dispersion relation of the system appears nicely in
Fig.~\ref{fig:figure10}(c).  The value of the gap at $q = \pi$ measured in
our simulation is $\Delta = 0.4105$, consistent with the value found
in Ref.~\onlinecite{White3}. Thus using the method we have presented here
we obtain a spectral function with comparable accuracy to previous calculations
but with significantly reduced computational effort.

\section{Conclusions}
We have introduced the infinite boundary condition as a procedure
for representing a finite section of a lattice embedded within
an infinite chain. With just two boundary
sites we can describe the relevant information for the whole
semi-infinite spin chain. Therefore, instead of simulating the iMPS,
we just need the finite MPS with two additional effective sites. This
helps to greatly reduce the computational cost as well as computer
memory in simulating the infinite system where the MPS cannot be
represented by translationally invariant tensors. After finding the
effective Hamiltonian and operators associated with the infinite
boundary, the numerical algorithms we use are straightforward,
variants of the well-known MPS/DMRG algorithms for finite-size systems.
Hence the general procedure is applicable to a wide variety of problems.

As an example for possible application we considered the real-time
evolution of the 1D spin-1 isotropic Heisenberg model. The initial
state of the system is the ground state where one central site is
locally perturbed. As a result, a wave packet is formed and spreads
out from the center in time. As we have already attached the infinite
boundaries to the finite system, we do not need to end the simulation
 when the wave front hits the boundaries, as useful information can be
still be obtained, at least for short time intervals, as the
degrees of freedom propagate into the boundary tensor.
The resulting spectral function and dispersion relation
compare well with previous investigations. The gap value we obtained
compares well with that obtained in Ref.~\onlinecite{White3}, although smaller
window size and longer evolution time are used. In fact, there is no restriction
in our method which says that the window size must remain constant throughout the
calculation. Expanding the window size is straightforward, as the tensors
representing the system outside the window are translationally invariant anyway,
so additional tensors
simply need to be orthogonalized and incorporated into the finite window. Similarly,
reducing the size of the window is achieved by incorporating tensors
from the finite window into the infinite boundary tensors, which is a simple
tensor contraction for the new effective Hamiltonian and associated operators.
The use of these techniques is described in Ref. \onlinecite{PhienMovingWindow}.

We have described the procedure for infinite boundary conditions for a one-dimensional matrix product state;
however the general procedure is applicable to any regular tensor network.
In particular, this method is directly applicable to iPEPS. \cite{Verstraete1,Murg1,Jordan1}
This may be an effective way to obtain the spectral function of a 2D system,
among many other possible applications.

\begin{acknowledgments}
we have recently learned of some related works.\cite{TomotoshiMovingWindow,OsborneMovingWindow}
We acknowledge support from the Australian
Research Council Centre of Excellence for Engineered
Quantum Systems and the Discovery Projects funding
scheme (Project No. DP1092513).
\end{acknowledgments}

\footnotesize

\end{document}